\documentclass[aps,english,preprint,pra]{revtex4-1}
\usepackage{amsmath}
\usepackage{SIunits}
\usepackage{graphicx}

\begin{document}
%
%
%
\newcommand{\ac}[0]{\ensuremath{\hat{a}_{\mathrm{c}}}}
\newcommand{\adagc}[0]{\ensuremath{\hat{a}^{\dagger}_{\mathrm{c}}}}
\newcommand{\aR}[0]{\ensuremath{\hat{a}_{\mathrm{R}}}}
\newcommand{\aT}[0]{\ensuremath{\hat{a}_{\mathrm{T}}}}
\renewcommand{\b}[0]{\ensuremath{\hat{b}}}
\newcommand{\bdag}[0]{\ensuremath{\hat{b}^{\dagger}}}
\newcommand{\betaI}[0]{\ensuremath{\beta_\mathrm{I}}}
\newcommand{\betaR}[0]{\ensuremath{\beta_\mathrm{R}}}
\renewcommand{\c}[0]{\ensuremath{\hat{c}}}
\newcommand{\cdag}[0]{\ensuremath{\hat{c}^{\dagger}}}
\newcommand{\CorrMat}[0]{\ensuremath{\boldsymbol\gamma}}
\newcommand{\Deltacs}[0]{\ensuremath{\Delta_{\mathrm{cs}}}}
\newcommand{\Deltacsmax}[0]{\ensuremath{\Delta_{\mathrm{cs}}^{\mathrm{max}}}}
\newcommand{\Deltacsparked}[0]{\ensuremath{\Delta_{\mathrm{cs}}^{\mathrm{p}}}}
\newcommand{\Deltacstarget}[0]{\ensuremath{\Delta_{\mathrm{cs}}^{\mathrm{t}}}}
\newcommand{\Deltae}[0]{\ensuremath{\Delta_{\mathrm{e}}}}
\newcommand{\Deltahfs}[0]{\ensuremath{\Delta_{\mathrm{hfs}}}}
\newcommand{\dens}[0]{\ensuremath{\hat{\rho}}}
\newcommand{\erfc}[0]{\ensuremath{\mathrm{erfc}}}
\newcommand{\Fq}[0]{\ensuremath{F_{\mathrm{q}}}}
\newcommand{\gammapar}[0]{\ensuremath{\gamma_{\parallel}}}
\newcommand{\gammaperp}[0]{\ensuremath{\gamma_{\perp}}}
\newcommand{\gavg}[0]{\ensuremath{\mathcal{G}_{\mathrm{avg}}}}
\newcommand{\gbar}[0]{\ensuremath{\bar{g}}}
\newcommand{\gens}[0]{\ensuremath{g_{\mathrm{ens}}}}
\renewcommand{\H}[0]{\ensuremath{\hat{H}}}
\renewcommand{\Im}[0]{\ensuremath{\mathrm{Im}}}
\newcommand{\kappac}[0]{\ensuremath{\kappa_{\mathrm{c}}}}
\newcommand{\kappamin}[0]{\ensuremath{\kappa_{\mathrm{min}}}}
\newcommand{\kappamax}[0]{\ensuremath{\kappa_{\mathrm{max}}}}
\newcommand{\ket}[1]{\ensuremath{|#1\rangle}}
\newcommand{\mat}[1]{\ensuremath{\mathbf{#1}}}
\newcommand{\mean}[1]{\ensuremath{\langle#1\rangle}}
\newcommand{\omegac}[0]{\ensuremath{\omega_{\mathrm{c}}}}
\newcommand{\omegas}[0]{\ensuremath{\omega_{\mathrm{s}}}}
\newcommand{\pauli}[0]{\ensuremath{\hat{\sigma}}}
\newcommand{\pexc}[0]{\ensuremath{p_{\mathrm{exc}}}}
\newcommand{\pexceff}[0]{\ensuremath{p_{\mathrm{exc}}^{\mathrm{eff}}}}
\newcommand{\Pa}[0]{\ensuremath{\hat{P}_{\mathrm{c}}}}
\newcommand{\Qmin}[0]{\ensuremath{Q_{\mathrm{min}}}}
\newcommand{\Qmax}[0]{\ensuremath{Q_{\mathrm{max}}}}
\renewcommand{\Re}[0]{\ensuremath{\mathrm{Re}}}
\renewcommand{\S}[0]{\ensuremath{\hat{S}}}
\newcommand{\Sminuseff}[0]{\ensuremath{\hat{S}_-^{\mathrm{eff}}}}
\newcommand{\Sxeff}[0]{\ensuremath{\hat{S}_x^{\mathrm{eff}}}}
\newcommand{\Syeff}[0]{\ensuremath{\hat{S}_y^{\mathrm{eff}}}}
\newcommand{\tildeac}[0]{\ensuremath{\tilde{a}_{\mathrm{c}}}}
\newcommand{\tildepauli}[0]{\ensuremath{\tilde{\sigma}}}
\newcommand{\Tcaveff}[0]{\ensuremath{T_{\mathrm{cav}}^{\mathrm{eff}}}}
\newcommand{\Techo}[0]{\ensuremath{T_{\mathrm{echo}}}}
\newcommand{\Tmem}[0]{\ensuremath{T_{\mathrm{mem}}}}
\newcommand{\Tswap}[0]{\ensuremath{T_{\mathrm{swap}}}}
\newcommand{\Var}[0]{\ensuremath{\mathrm{Var}}}
\renewcommand{\vec}[1]{\ensuremath{\mathbf{#1}}}
\newcommand{\Xa}[0]{\ensuremath{\hat{X}_{\mathrm{c}}}}

\title{Towards a spin-ensemble quantum memory for superconducting qubits}

\author{C. Grezes$^{1}$, Y. Kubo$^{1}$, B. Julsgaard$^{2}$, T. Umeda$^{3}$, J. Isoya$^{4}$, H. Sumiya$^{5}$, H. Abe$^{6}$, S. Onoda$^{6}$, T. Ohshima$^{6}$, K. Nakamura$^{7}$, I. Diniz$^{8}$, A. Auffeves$^{8}$, V. Jacques$^{9,10}$, J.-F. Roch$^{10}$, D. Vion$^{1}$, D. Esteve$^{1}$, K. M{\o}lmer$^{2}$, and P. Bertet$^{1}$}

\affiliation{$^{1}$Quantronics group, Service de Physique de l'Etat Condens{\'e}, DSM/IRAMIS/SPEC, CNRS UMR 3680, CEA Saclay,
91191 Gif-sur-Yvette, France}

\affiliation{$^{2}$Department of Physics and Astronomy, Aarhus University, Ny Munkegade 120, DK-8000 Aarhus C, Denmark.}

\affiliation{$^{3}$Institute of Applied Physics, University of Tsukuba, Tsukuba 305-8573, Japan}

\affiliation{$^{4}$Research Center for Knowledge Communities, University of
Tsukuba, Tsukuba 305-8550, Japan}

\affiliation{$^{5}$Sumitomo Electric Industries Ltd., Itami 664-001, Japan}

\affiliation{$^{6}$Japan Atomic Energy Agency, Takasaki 370-1292, Japan}

\affiliation{$^{7}$Energy System Research Institute, Fundamental Technology Department, Tokyo Gas Co., Ltd., Yokohama, 230-0045, Japan}

\affiliation{$^{8}$Institut N\'eel, CNRS, BP 166, 38042 Grenoble, France}

\affiliation{$^{9}$Laboratoire Charles Coulomb, Université de Montpellier and CNRS, 34095 Montpellier, France}

\affiliation{$^{10}$Laboratoire Aim{\'e} Cotton, CNRS, Université Paris-Sud and ENS Cachan, 91405 Orsay, France}


\date{\today}

\begin{abstract}
This article reviews efforts to build a new type of quantum device, which combines an ensemble of electronic spins with long coherence times, and a small-scale superconducting quantum processor. The goal is to store over long times arbitrary qubit states in orthogonal collective modes of the spin-ensemble, and to retrieve them on-demand. We first present the protocol devised for such a multi-mode quantum memory. We then describe a series of experimental results using NV center spins in diamond, which demonstrate its main building blocks: the transfer of arbitrary quantum states from a qubit into the spin ensemble, and the multi-mode retrieval of classical microwave pulses down to the single-photon level with a Hahn-echo like sequence. A reset of the spin memory is implemented in-between two successive sequences using optical repumping of the spins.
\vspace{1cm}
\end{abstract}
\maketitle

Superconducting qubits are attractive candidates for quantum information processing because of their flexibility and rapid single- and two-qubit gates~\cite{Barends.Phys.Rev.Lett.111.080502(2013)}, but compared to microscopic systems they suffer from short coherence times. This suggests to combine the two types of systems in a hybrid quantum device, where microscopic entities with long coherence times act as a quantum memory to complement a small-scale quantum processor made of superconducting qubits~\cite{Xiang.RevModPhys.85.623(2013),Kurizki.PNAS.112.3866(2015)}. To be truly useful in a quantum computer, a quantum memory should behave as an ideal $n$-qubit register, with $n \gg 1$. It should be possible to re-initialize the memory (reset step), to transfer arbitrary qubit states $\ket{\psi_i}$ into each register (write step), store them over long times, and finally transfer back on-demand one of the states into the processor when needed for running the algorithm, while keeping the others in the memory (read step).

\begin{figure}[t!]
  \centering
  \includegraphics[width=12cm]{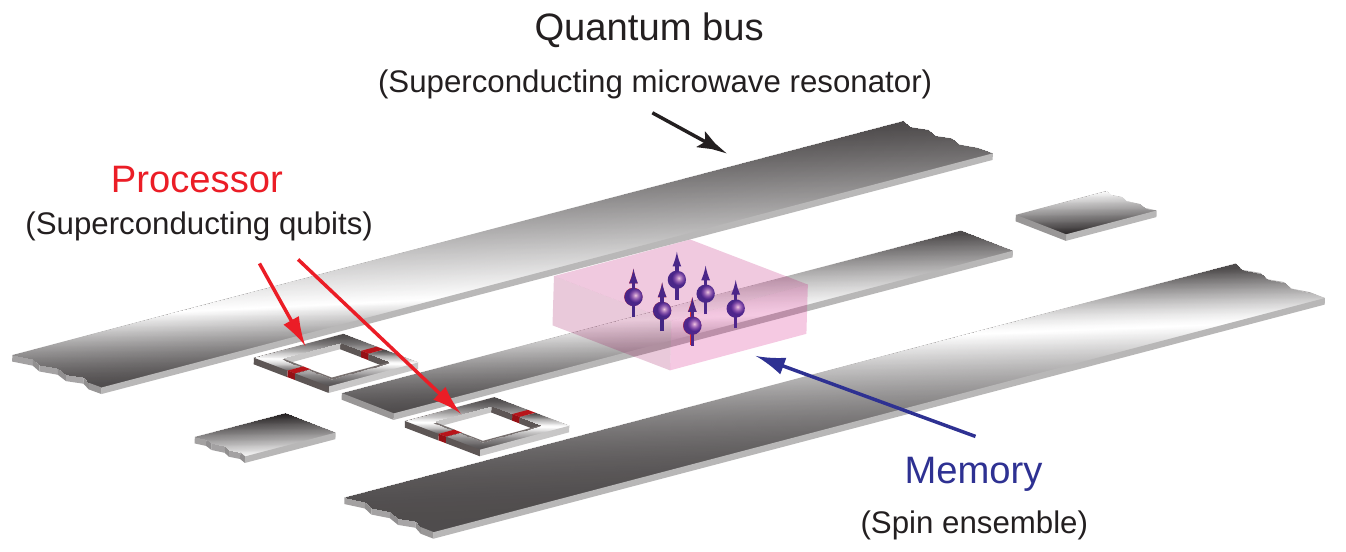}
  \caption{Concept of a hybrid quantum processor combining a two-qubit processor
and a spin ensemble multimode quantum memory. The exchange of quantum states between the processor and the memory is implemented by a superconducting microwave resonator used as a quantum bus, which is
coupled electrically to the qubits and magnetically to the spins.}
	\label{fig:HybridsQMPrinciple}
\end{figure}

Spins in crystals are ideally suited to implement these ideas, as already demonstrated in optical quantum memories~\cite{Lvovsky.NaturePhotonics3.706-714(2009)}. Large ensembles of $N$ spins offer as many orthogonal degrees of freedom, which can be exploited to store many qubit states in parallel. Various different systems are being investigated for this purpose: rare-earth ions in silicate crystals~\cite{Afzelius.NJP.1367-2630-15-6-065008(2013)}, donor spins in silicon~\cite{Wu.PhysRevLett.105.140503(2010)}, ... In this work we will focus on Nitrogen-Vacancy (NV) color centers in diamond as the storage medium. NV centers have coherence times that can reach seconds~\cite{Bargill.NatCom.4.1743(2013)} and can be actively reset in the spin ground state by optical repumping~\cite{Manson.PhysRevB.74.104303(2006)}. In our quantum memory project, the interaction between the qubits and the NV ensemble is mediated by a superconducting resonator used as a quantum bus, which is electrically coupled to the qubits, and magnetically coupled to the spins (see Fig.~\ref{fig:HybridsQMPrinciple}). The goal of this article is to review the experimental progress towards the implementation of the memory protocol~\cite{Julsgaard.PhysRevLett.110.250503} described below.

\section{Quantum memory protocol}
\label{sec:protocol}

Our memory protocol~\cite{Julsgaard.PhysRevLett.110.250503} is inspired by related work on optical quantum memories~\cite{Damon.NewJPhys.13.093031(2011),McAuslan.PhysRevA.84.022309(2011)}, adapted to the requirements of a circuit quantum electrodynamics (cQED) setup (see also~\cite{Wesenberg.PhysRevLett.103.070502(2009),Afzelius.NJP.1367-2630-15-6-065008(2013)}), which implies in particular working at millikelvin temperatures and microwave frequencies. As shown in Fig.~\ref{fig:HybridsQMPrinciple}, the qubit and spin ensemble are embedded inside the quantum bus microwave resonator. From numerous experiments in circuit QED (see for instance~\cite{hofheinz_synthesizing_2009}), it is well known that the state of a qubit can be transferred with high fidelity into the photonic state of the resonator; we will therefore exclusively focus on the issue of quantum state transfer between the resonator and the spin ensemble. We thus assume that the resonator receives a well-defined state $\ket{\psi} = \alpha \ket{0} + \beta \ket{1}$ (with $\ket{n}$ the n-photon Fock state) from the quantum processor (QuP), and our goal is to store this state into the long-lived spin ensemble. For simplicity, we will focus here on the case of a single-photon Fock state as input state $\ket{\psi} = \ket{1}$ - since the protocol is phase-preserving, a successful storage of $\ket{1}$ implies the same for arbitrary states. We will use the possibility offered in circuit QED to design and operate resonators whose frequency~\cite{PalaciosLaloy2008,sandberg_tuning_2008} and quality factor~\cite{Yin.PhysRevLett.110.107001(2013),Wenner.PhysRevLett.112.210501(2014)} can be dynamically tuned by inserting SQUIDs in the resonator or in the coupling circuit to the measurement line.

It is useful here to introduce a few elements of modelling for the coupling of an ensemble of $N$ emitters (the spins in our case), of frequency $\omega_j$, to a single resonator mode, of frequency $\omega_r$, with a coupling constant $g_j$, described by the Hamiltonian $H/\hbar = \omega_r a^\dagger a + \sum_j (\omega_j / 2) \sigma_{z,j} + \sum g_j(a\sigma_{+,j} + a^\dagger \sigma_{-,j})$, $a$ being the resonator annihilation operator, and $\sigma_j$ the Pauli operators associated to spin $j$. As long as the number of excitations is small compared to $N$, it is possible to replace the spin operators $\sigma_{-,j}$ by bosonic operators $s_j$. Introducing the {\it bright} mode operator $b = (\sum g_j s_j)/g_{ens}$, with $g_{ens} = \sqrt{ \sum g_j^2}$, the Hamiltonian can be rewritten as $H/\hbar = \omega_r a^\dagger a + \sum_j \omega_j s_j^\dagger s_j + g_{ens}(ab^\dagger  + a^\dagger b)$. Denoting $\ket{G}$ the collective ground state of the ensemble, and $\ket{E} = b^\dagger \ket{G}$ the first excited state of the bright mode, one sees that the state $\ket{1,G}$ ($1$ photon in the resonator) is coupled to the state $\ket{0,E}$ with a strength $g_{ens}$ collectively enhanced by a factor $\sqrt{N}$ compared to the coupling of a single spin. In the presence of inhomogeneous broadening, characterized by the width $\Gamma$ of the distribution of resonance frequencies $\rho(\omega) = \sum |g_j|^2 \delta (\omega - \omega_j)$, the state $\ket{0,E}$ is not an energy eigenstate and hence is not stationary. Instead it decays towards the other $N-1$ single-excitation states of the spin-ensemble manifold (called {\it dark} modes because they are de-coupled from the radiation field) in a characteristic time called the Free-Induction Decay (FID) time $T_2^* = \Gamma^{-1}$~\cite{Dicke,Kubo.PhysRevA.85.012333}.

\begin{figure}[t!]
  \centering
  \includegraphics[width=16cm]{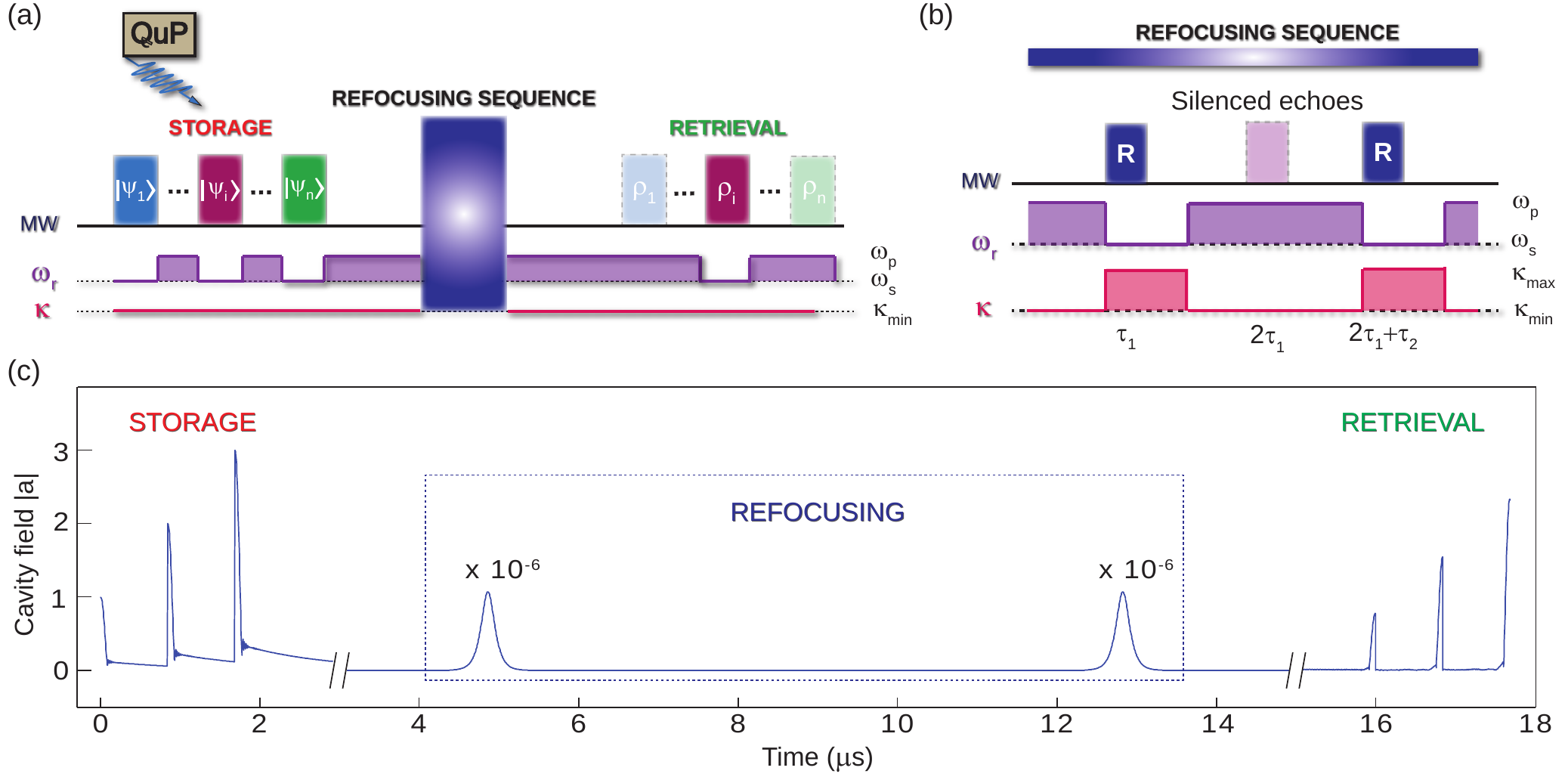}
  \caption{Quantum memory protocol. (a) Schematic timing of microwave pulses and cavity parameters $\omega_r$ and $\kappa$. Quantum states $\ket{\psi_i}$ delivered by the superconducting qubits are stored in the spin ensemble (write step). A refocusing sequence acts as  time-reversal for the spins and triggers the retrieval of the stored states as an echo (read step). On-demand retrieval is obtained by dynamically detuning the cavity to retrieve a state $\ket{\psi_i}$. (b) Refocusing sequence. The refocusing sequence involves two refocusing pulse bringing back the spin ensemble in the original first excited state of the bright mode  (with most of the spins in their ground state). The first echo is "silenced" to preserve the information for the second echo. The stabilization of the inverted spin ensemble is ensured by setting $\kappa=\kappa_{max}$ during and immediately after the refocusing pulses. (c) Results of numerical simulations, showing the cavity field versus time in a multi-mode storage example with three input fields separated by $0.84$\,$\mu$s and memory time $16$\,$\mu$s.}
	\label{fig:ProtocolFig}
\end{figure}

The quantum memory protocol proceeds in two distinct steps (see Fig.~\ref{fig:ProtocolFig}a). In the first step ({\it write}), the spin ensemble and resonator are tuned into resonance. If the collective coupling strength satisfies the strong coupling condition $g_{ens} \gg \kappa,\Gamma$, with $\kappa = \omega_r / Q$ the resonator energy decay rate and $Q$ its quality factor, the microwave photon in the cavity is coherently absorbed by the spin ensemble as an elementary excitation of the bright mode $\ket{E}$. After a time of the order of a few $T_2^*$, the bright mode excitation is redistributed over the dark modes, a new incoming quantum state can be written into the bright state mode. In this way, a large number of quantum states can be sequentially written into the orthogonal modes of the ensemble. 

The second step ({\it read}) consists in recovering on-demand one desired state out of the ensemble (without perturbing the storage of all the others). This is achieved using a refocusing sequence (see Fig.~\ref{fig:ProtocolFig}b), based on the principle of the Hahn echo~\cite{Hahn.PhysRev.80.580(1950)}: the FID of a transverse magnetization created at time $t=0$ by a microwave pulse, due to the spread in resonance frequency of the spins, can be reversed by applying a $\pi$ pulse to the spins at a later time $\tau$ that effectively acts as a time reversal and causes the spins to return in phase at time $2 \tau$, leading to the emission of an echo with a well-defined phase and amplitude relation with the initial pulse. This sole idea is however not sufficient for retrieving a single photon with high fidelity. Indeed an echo generated in this way cannot reproduce faithfully the quantum statistics of the initial microwave field state because the $\pi$ pulse inverts the spin excitation, and an inverted spin ensemble is an amplifying medium which unavoidably will add extra noise to the emitted photon~\cite{Chaneliere.PhysRevA.79.053851(2009)}. 

In order to reach the quantum regime, one possibility is to use a sequence of two $\pi$ pulses, in which case the initial quantum states can indeed be recovered with high-fidelity within the second echo since the spin ensemble is then no longer inverted. However this requires actively suppressing the leakage of the stored quantum information during the first echo. In optical quantum memories, such an echo silencing can be performed by applying reversible electric field gradients~\cite{McAuslan.PhysRevA.84.022309(2011)}, or using propagation effects~\cite{Damon.NewJPhys.13.093031(2011)}. In cQED, one can instead detune the resonator frequency $\omega_r$ from the spin frequency $\omega_s$ during the first echo to avoid emission, and re-tune it in resonance with the spins only at the time at which the second echo is expected~\cite{Julsgaard.PhysRevLett.110.250503,Afzelius.NJP.1367-2630-15-6-065008(2013),Julsgaard.PhysRevA.88.062324(2013)}. Note that this idea is straightforwardly extended to the case where several quantum states are stored sequentially in the ensemble; tuning $\omega_r$ to $\omega_s$ only at the time of the echo $i$ allows then an on-demand retrieval of state $\ket{\psi_{i}}$ while keeping the other states unperturbed in the memory. Silencing the first echo is actually not entirely sufficient to solve the issues linked to the inversion of the spin ensemble by the first $\pi$-pulse. Another threat arises from the fact that an inverted spin ensemble can desexcite itself much faster than the spontaneous emission rate of an individual spin, a phenomenon known as superradiance. Superradiance occurs only if the strong coupling criterion is satisfied~\cite{Julsgaard.PhysRevA.86.063810(2012)}; therefore, during all the time that the spin ensemble is inverted the system should be in the weak coupling regime in order to stabilize the ensemble~. In our protocol we propose to achieve this by increasing dynamically the cavity damping rate $\kappa$; this makes it possible to switch between the strong and weak coupling regimes in the course of the experiment, as required for high-fidelity {\it write} and {\it read} steps.

The storage time of such a quantum memory is ultimately limited by the time $T_2$ at which the spin coherence is lost. Since the write step duration for a given quantum state is $\approx T_2^*$, one sees that approximately $T_2 / T_2^*$ quantum states can be separately stored in the memory. This number can reach $10^2 - 10^3$ in many systems. The fidelity of this protocol was calculated numerically in~\cite{Julsgaard.PhysRevLett.110.250503}, and was shown to reach up to $80\%$ using realistic values for the physical parameters (see Fig.~\ref{fig:ProtocolFig}c).

\section{NV centers in diamond}
\label{sec:NV}

\begin{figure}[t!]
  \centering
  \includegraphics[width=12cm]{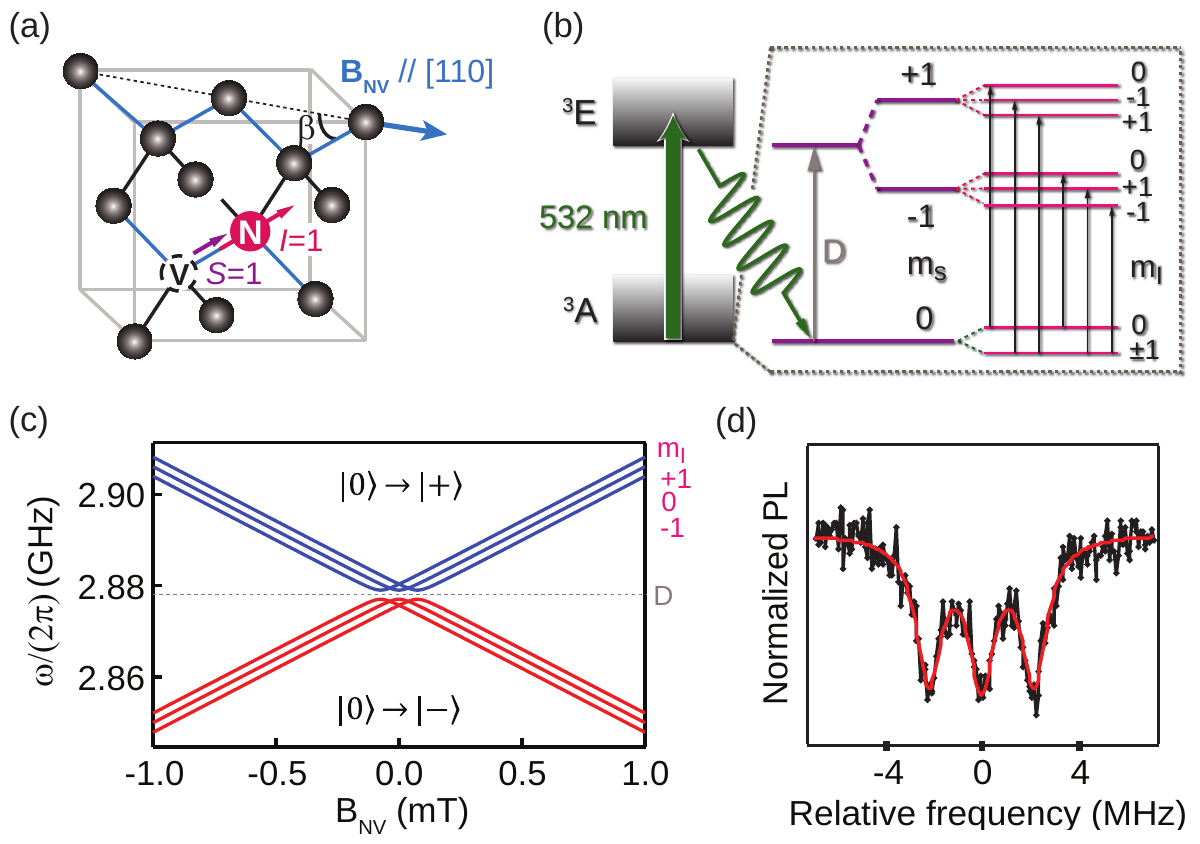}
  \caption{NV centers in diamond. (a) Negatively-charged NV centers consist of a substitutional Nitrogen atom next to a Vacancy of the diamond lattice, having trapped an electron. The electronic spin $S=1$ is coupled to the $I=1$ nuclear spin of the nitrogen atom by hyperfine interaction. A magnetic field $B_{NV}$ is applied to lift the degeneracy between transitions. (b) NV simplified energy diagram showing the ground ${}^3$A and the excited ${}^3$E electronic states as well as the Zeeman and hyperfine structure of ${}^3$A, with a zero-field splitting $D/2\pi=2.878$\,GHz. NV centers can be excited optically in the excited orbital state, through which they are repumped into the $m_S=0$ state. (c) Transition frequency from the spin ground state $\ket{m_S=0,m_I}$ to the excited states $\ket{\pm,m_I}$, in the case of a NV center having an axis along the $(111)$ crystalline direction, and with $B_{NV}$ parallel to $(110)$. (d) Optically-Detected electron spin resonance (ESR) hyperfine spectrum of an ensemble of NV centers in a diamond crystal used for the experiment reported in paragraph~\ref{sec:NVQubit} and in~\cite{Kubo.PhysRevLett.107.220501(2011)}.}
	\label{fig:HybridsNV}
	\end{figure}

NV centers are diamond defects consisting of a substitutional nitrogen atom (N) sitting next to a vacancy (V) of the diamond lattice (see Fig.~\ref{fig:HybridsNV}a). In their negatively-charged state ($NV^-$), NV centers have very interesting electronic properties arising from their peculiar energy level scheme~\cite{GruberScience276.2012(1997)}. First, the electronic ground state of the NV is a spin-triplet $S=1$, with a zero-field splitting $D/2\pi = 2.88$\,GHz between states $m_S = \pm 1 $ and $m_S = 0$, quantized along the NV center symmetry axis, arising from spin-spin interactions in the diamond crystal field. The complete NV ground state spin Hamiltonian needs to be presented in some detail, since this is the degree of freedom that is used for storing the quantum information in our experiments. Physically, four terms contribute to this Hamiltonian : (i) the zero-field splitting, (ii) the Zeeman energy shift in an external magnetic field $\overrightarrow{B}_{NV}$, (iii) the interaction of the NV spin with local electric field or strain in the diamond due to the spin-orbit coupling, which causes the states $m_S = \pm 1$ to hybridize around zero magnetic field, and (iv) the hyperfine (HF) interaction of the NV electronic spin with the nuclear spin of the nitrogen atom, which is also a spin triplet $I=1$ for the most common isotope $^{14}N$ used in our experiments. In total, the spin Hamiltonian~\cite{Neumann2009} writes

\begin{equation}
H_{NV} / \hbar = D S_z^2 + \gamma_e \overrightarrow{B_{NV}} \cdot \overrightarrow{S} + E (S_x^2 - S_y^2) + A_z S_z I_z + \mathcal{Q} [I_z^2 - I(I+1)/3],
\end{equation}

\noindent with $S_{x,y,z}$ (resp. $I_{x,y,z}$) the spin angular momentum operators of the NV electron (resp. nitrogen nucleus), $E$ the coupling induced by local strain and electric fields, $A_z / 2\pi =- 2.1$\,MHz, $\gamma_e/2\pi = -28$\,GHz/T the electronic gyromagnetic ratio, and $\mathcal{Q}/2\pi=-5$\,MHz the quadrupolar momentum of the nitrogen nuclear spin. The resulting energy levels are shown schematically in Fig.~\ref{fig:HybridsNV}b. Only transitions between levels with the same nuclear spin state $m_I$ are allowed, and their frequency from the ground state $\ket{m_S=0, m_I}$ to $\ket{\pm, m_I}$ is shown in Fig.~\ref{fig:HybridsNV}c in the case where $E/2\pi = 2$\,MHz. Note that because of the spin-orbit term, the energy eigenstates $\ket{\pm}$ correspond to the pure spin states $m_S = \pm 1$ only for large magnetic fields, and are linear combinations of these spin states around zero magnetic field. From Fig.~\ref{fig:HybridsNV}c one sees that superconducting circuits with resonance frequencies around $2.9$\,GHz can be brought into resonance with one of these spin transitions in a magnetic field as small as $1$\,mT, which is important since superconducting circuits have increased dissipation in large magnetic fields due to magnetic vortices.

Another aspect is the optical properties of NV centers : when excited with green light, they fluoresce in the red. This is due to their excited electronic state (see Fig.~\ref{fig:HybridsNV}b), which is also a spin-triplet. Ground-state to excited state transitions conserve the electronic spin; however the NV relaxation from its excited state is spin-dependent: the $m_S = \pm 1$ state has a large probability of relaxing via a singlet metastable state (not shown in the energy diagram of Fig.~\ref{fig:HybridsNV}b), which then relaxes to $m_S=0$ in the ground state. After a few optical cycles, the NV center spin state is therefore repumped in $m_S=0$ with a $\sim 90\%$ efficiency~\cite{Robledo.Nature.477.574(2011)}. Since the metastable state is long-lived, the fluorescence intensity is reduced if the NV spin is initially in $m_S = \pm 1$ compared to $m_S=0$. These remarkable properties enable the optical detection of the spin state of indivual NV centers (so-called Optical Detection of Magnetic Resonance ODMR) at room-temperature, using a straightforward confocal microscope setup. For instance the spectroscopy of an ensemble of NV centers, measured at room-temperature by ODMR on one of our diamond samples, is shown in Fig.~\ref{fig:HybridsNV}d. It displays the typical hyperfine structure of the $^{14}$NV center transitions with a triplet of lines separated by $2.17$\,MHz.

The interest of using NV centers for storing quantum information stems from the fact that long coherence times have been observed on the ground state spin transition in ultra-pure crystals. The coherence properties of NV centers are thus of utter importance. They are characterized by the Free-Induction Decay time $T_2^*$ (measured by Ramsey fringes), the Hahn-echo decay time $T_2$ (measured by a spin-echo sequence), and the coherence time under dynamical decoupling sequences such as Carr-Purcell-Meiboom-Gil $T_{2CPMG}$. It is well-established that the values found for all these times depend crucially on the local magnetic environment of each NV center~\cite{vanWyk.0022-3727-30-12-016(1997),Zhao.PhysRevB.85.115303(2012)}. In diamond, the main magnetic impurities surrounding the NVs are either neutral nitrogen atoms (so-called $P1$ centers) which are paramagnetic with an electron spin $1/2$, or spin 1/2 carbon $13$ nuclei present to $1.1\%$ abundance in natural carbon. The longest coherence times were therefore measured in ultra-pure samples grown by Chemical-Vapor Deposition (CVD) with very low nitrogen concentration as well as an isotopically enriched carbon source. In such samples, $T_2^*=500 \mu \mathrm{s}$~\cite{Maurer.Science.336.1283(2012)} and $T_2 = 2$\,ms~\cite{Balasubramian.NatMat.8.383(2009)} have been reported at room-temperature; at lower temperatures ($100$\,K), $T_{2CPMG} = 0.5$\,s was reached~\cite{Bargill.NatCom.4.1743(2013)}.

We need relatively large concentrations of NV centers of order $\sim 1-10$\,ppm to efficiently absorb the microwave radiation. These concentrations are not easily reached with samples grown by CVD, and our crystals are grown by another method called High-Pressure-High-Temperature (HPHT). HPHT diamonds usually have a large nitrogen concentration of $1-100$\,ppm. To create NV centers the crystals are irradiated (either with protons or electrons) to create vacancies in the lattice; they are then annealed at $800-1000\,^{\circ}{\rm C}$ for few hours so that the vacancies migrate and form NV centers when they meet a nitrogen atom. This method unavoidably leaves a significant residual concentration of P1 centers ($ 1-100$\,ppm), which limits the spin coherence time (both $T_2^*$ and $T_2$) to lower values than reported above. In most experiments reviewed here (except in the last results of section~\ref{sec:StorageRetrieval}), these residual P1 centers are the main cause of decoherence of the NV centers.

\section{Strong coupling between a spin ensemble and the resonator}
\label{sec:strongcoupling}

As explained in the description of the quantum memory protocol, the {\it write} step requires to reach the strong coupling regime between the spin ensemble and the resonator, i.e. $g_{ens} \gg \kappa, \Gamma$. The Hamiltonian of a single NV center coupled to the electromagnetic field in a resonator is $H = -  \gamma_e \overrightarrow{S} \cdot \overrightarrow{B}$ and $\overrightarrow{B} = \overrightarrow{\delta B_0}(a + a^\dagger)$, $|\overrightarrow{\delta B_0}(\overrightarrow{r})|$ being the rms vacuum fluctuations of the magnetic field in the resonator mode at the spin position $\overrightarrow{r}$. Restricting this Hamiltonian to one of the NV transitions (for instance the $m_S=0 \rightarrow m_S=+1$ transition\footnote{We suppose here to be in the high field limit where the energy eigenstates correspond to the pure spin states}), it can be rewritten in a Jaynes-Cummings form :

\begin{equation}
H = \hbar g(\overrightarrow{r}) (\sigma_+ a + \sigma_- a^\dagger),
\end{equation}
\noindent
with $\hbar g(\overrightarrow{r})= \gamma_e \langle m_S = +1 |\overrightarrow{S} | m_S = 0 \rangle \cdot \overrightarrow{\delta B_0}(\overrightarrow{r}) $. If $\theta(\overrightarrow{r})$ is the angle between the NV axis and the local microwave field, choosing $z$ as the NV axis and $x$ as the orthogonal direction in the $(z,\overrightarrow{\delta B_0}(\overrightarrow{r}))$ plane, one finds $ \hbar g= \gamma_e |\langle m_S = +1 |S_x | m_S = 0 \rangle| |\delta B_0(\overrightarrow{r})| |\sin \theta(\overrightarrow{r})|$, yielding

\begin{equation}
g(\overrightarrow{r}) = \gamma_e |\delta B_0(\overrightarrow{r})| |\sin \theta(\overrightarrow{r})| / \sqrt{2},
\end{equation}
\noindent
since $\langle m_S = +1 |S_x | m_S = 0 \rangle = \hbar / \sqrt{2}$. The coupling constant is thus governed by the amplitude of the vacuum field fluctuations at the spin location $|\delta B_0(\overrightarrow{r})|$. For a conventional CPW resonator as frequently used in cQED (with transverse dimensions of order $10 \mu \mathrm{m}$ and a $50\Omega$ characteristic impedance), one finds $|\delta B_0| \approx 0.4$\,nT, yielding a coupling constant $g/2\pi \approx 10$\,Hz for $\theta = \pi / 2$.

The ensemble coupling constant is then obtained as $g_{ens} = [\int g^2(r) \rho(r)]^{1/2} $, $\rho(r)$ being the local NV density. For a sample of volume $V$ with a homogeneous NV center concentration $\rho$, one gets that $g_{ens} = |\gamma_e| \sqrt{\frac{\mu_0 \hbar \omega_r \rho \alpha \eta}{4}}$, $\eta =  \int_V d \mathbf{r} |\delta B_{\mathrm{0}}(\mathbf{r}) |^2 / \int d \mathbf{r} |\delta B_{\mathrm{0}}(\mathbf{r})|^2$ being the so-called mode filling factor and $\alpha =  \int_V d \mathbf{r} |\delta B_{\mathrm{0}}(\mathbf{r}) \sin \theta(\mathbf{r}) |^2 / \int_V d \mathbf{r} |\delta B_{\mathrm{0}}(\mathbf{r}) |^2$ a numerical factor depending on the orientation of the crystal relative to the resonator, with $\alpha$ and $\eta$ verifying $0<\alpha,\eta<1$\cite{Kubo.PhysRevLett.105.140502(2010),Grezes.PhD(2014)}. Reaching the strong coupling regime requires therefore samples with large concentration $\rho$, maximum filling factor, and narrow linewidth $\Gamma$. This regime is manifested by the appearance of an avoided level crossing in the resonator spectrum, indicating the hybridization between the resonator and the spin ensemble. Such a phenomenon was observed in $2010$ with an ensemble of NV centers coupled to a half-wavelength coplanar-waveguide resonator whose frequency $\omega_r(\Phi_x)$ was made tunable using a SQUID-array~\cite{Kubo.PhysRevLett.105.140502(2010)}.

\begin{figure}[t!]
  \centering
  \includegraphics[width=6cm]{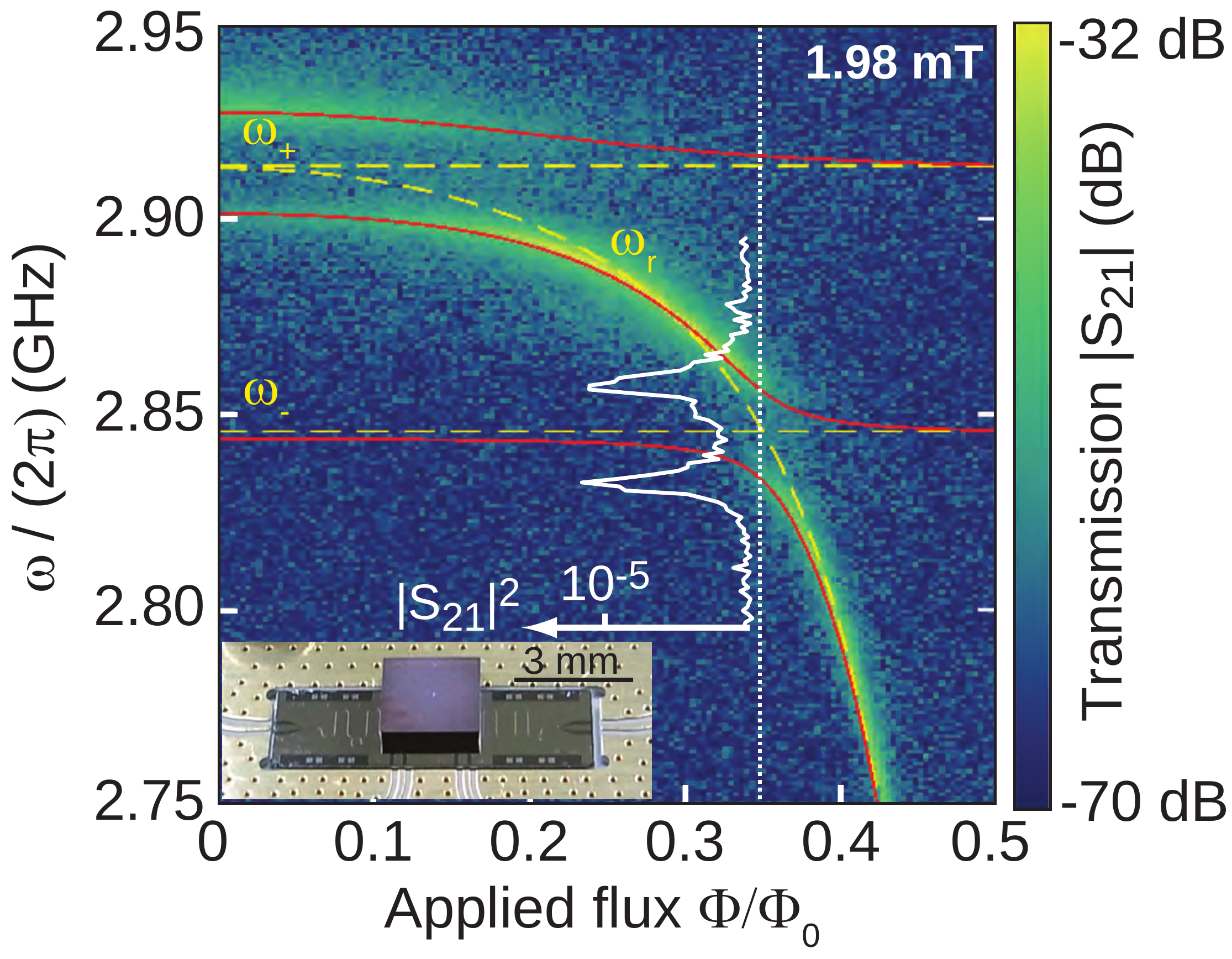}
  \caption{Spectroscopic evidence for spin ensemble-resonator strong coupling. The resonator transmission spectrum $|S_{21}|$ shows two anticrossings when the tunable resonator frequency crosses the NV transitions, with a field $B_{NV} = 1.98$\,mT applied along the $(100)$ crystalline direction. Red solid (yellow dashed) lines are fits to the eigenfrequencies of the coupled (decoupled) resonator-spin system. A transmission spectrum (white overlay) is also shown in linear units in the middle of the anticrossing with the NV transition $\ket{0} \rightarrow \ket{-}$ (extracted from~\cite{Kubo.PhysRevLett.105.140502(2010)}).}
	\label{fig:HybridsStrongCouplingSpectro}
	\end{figure}

In the experiment, the diamond crystal (of the HPHT type, with $\rho \approx 10^{6} \mu \mathrm{m}^{-3}$ corresponding to $6$\,ppm, and a P1 center concentration of $\approx 150$\,ppm) is directly glued on top of the resonator, and positioned in the middle to maximize the filling factor.  The spin Zeeman splitting can be tuned with a magnetic field $\overrightarrow{B}_{NV}$ parallel to the sample surface along the $[100]$ axis, so that the four possible NV orientations within the crystal experience the same Zeeman shift. Only two spin resonance frequencies $\omega_\pm$ are therefore expected for the $\ket{0} \rightarrow \ket{\pm}$ transitions. Measurements of the resonator transmission $|S_{21}|(\omega)$ at $40$\,mK with the diamond crystal on top of the resonator circuit are shown in Fig.~\ref{fig:HybridsStrongCouplingSpectro}. Two-dimensional plots of the transmission spectrum as a function of $\Phi_x$ are presented for $B_{NV}=2$\,mT. Two avoided crossings are observed when the resonator is tuned through the NV center ESR frequencies, which reveals the strong coupling of the NV ensemble to the resonator. These results are modelled with a $3$-mode Hamiltonian $H/\hbar = \omega_r(\Phi_x) a^\dagger a + \omega_+ b_+ ^\dagger b_+ + \omega_- b_- ^\dagger b_- + g_+ (a^\dagger b_+ + h.c.) + g_- (a^\dagger b_- + h.c.)$, where two bright modes $b_\pm$ are introduced for each ESR frequency, as well as the corresponding collective coupling constants $g_{\pm}$. The transmission spectrum was fitted for each $\Phi_x$ by a sum of Lorentzian peaks whose central frequencies are then fitted to the eigenfrequencies of $H$, yielding the red solid lines in Fig.~\ref{fig:HybridsStrongCouplingSpectro}. The fitted coupling constants are $g_+ = g_- =2\pi \times 11$\,MHz, in quantitative agreement with the values estimated from the measured $\rho$. Since $\kappa = 9 \times 10^6 \mathrm{s}^{-1}$ and $\Gamma = 2 \pi \times 3$\,MHz, the strong coupling regime is indeed reached~\cite{Kubo.PhysRevLett.105.140502(2010)}. Since these measurements, normal mode splittings between a spin ensemble and a superconducting resonator mode have been observed in several different experiments: NV centers ~\cite{Amsuss.PhysRevLett.107.060502(2011)} and P1 centers~\cite{Schuster.PhysRevLett.105.140501(2010),Ranjan.PhysRevLett.110.067004} in diamond, rare-earth ions in $\mathrm{Y}_2\mathrm{SO}_4$~\cite{Probst.PhysRevLett.110.157001(2013)}, and magnonic excitations in YIG~\cite{Huebl.PhysRevLett.111.127003(2013),Tabuchi.PhysRevLett.113.083603(2014)}.

\section{Transfer of a qubit state into the NV ensemble}
\label{sec:NVQubit}

\begin{figure}[t!]
  \centering
  \includegraphics[width=16cm]{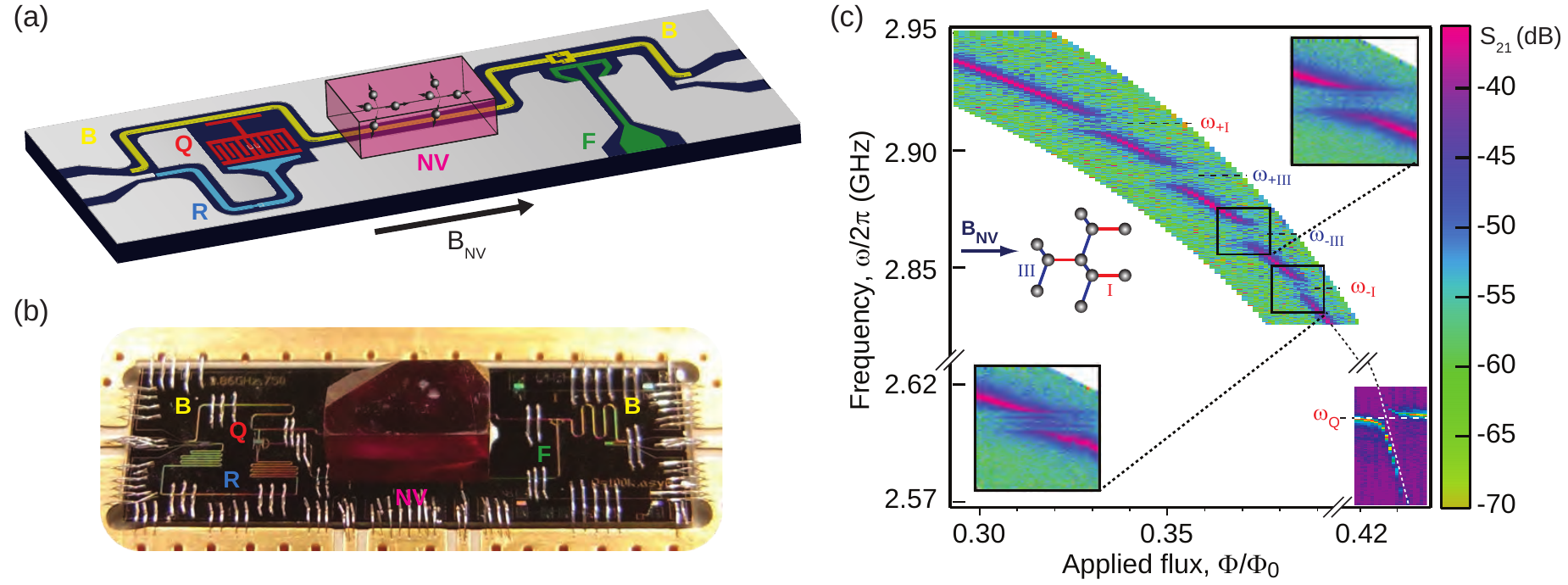}
  \caption{Storage of a qubit state: Sample layout. (a) Schematic view of the hybrid circuit, including a diamond crystal with the NV ensemble $NV$, and a transmon qubit $Q$ (in red). They are coupled via the quantum bus coplanar resonator $B$ (in yellow) whose frequency can be adjusted by passing current in a flux line (in green). The qubit is driven and readout via the bistable resonator $R$ (in blue). The magnetic field $B_{NV}$ is applied parallel to the sample surface, along $(111)$. (b) Photograph of the sample with diamond on top (the purple color is due to the NV centers). (c) Transmission coefficient amplitude $|S_{21}|$ of $B$, showing $4$ anticrossings corresponding to the $2$ ESR frequencies $\omega_{\pm}$ of each of the $2$ NV families $I, III$ undergoing different Zeeman shift due to their crystalline orientation as shown in the right top inset. At lower frequency an anti-crossing with the transmon qubit is seen at $\omega_Q / 2\pi = 2.607$\,GHz (extracted from~\cite{Kubo.PhysRevLett.107.220501(2011)}).}
	\label{fig:NVQubit1}
	\end{figure}

After the spectroscopic evidence for strong coupling between an ensemble of NVs and a resonator, the next step was to add a superconducting qubit on the chip with its readout circuit and to demonstrate the transfer of its quantum state into the NV spin ensemble. A schematic picture of the experiment is shown in Fig.~\ref{fig:NVQubit1}. We integrate on the same chip the diamond crystal containing the NV center ensemble (denoted $\mathrm{NV}$) and a transmon qubit $\mathrm{Q}$ (with transition frequency $\omega_\mathrm{Q}$ between its levels \ket{g} and \ket{e}). Their interaction is mediated by a quantum bus: a coplanar resonator $\mathrm{B}$ whose frequency $\omega_\mathrm{B}$ can be tuned on a nanosecond timescale by changing the flux $\Phi$ through the loop of a SQUID integrated in the resonator central conductor. The qubit is driven and readout through a second dedicated resonator $\mathrm{R}$, made non-linear by inserting a Josephson junction in its middle, which transforms the resonator into a high-fidelity sample-and-hold detector of the qubit state~\cite{mallet_single-shot_2009} enabling a direct measurement of the qubit excited state probability $P_e$.

The diamond is a HPHT crystal with $40$\,ppm initial nitrogen concentration (i.e. P1 centers). After irradiation and annealing, the NV concentration measured by photoluminescence is $2.5$\,ppm, yielding a final P1 concentration of $\approx 35$\,ppm, i.e. $3-4$ times lower than in the experiment reported above. As a result, the NV linewidth is also narrower, as can be seen from the ODMR spectrum shown in Fig.~\ref{fig:HybridsNV} which was measured on this sample, where the hyperfine NV structure is well-resolved with a linewidth of $\approx 1.5$\,MHz for each peak.

As in the previous experiment, the diamond crystal is glued on top of the resonator with vacuum grease. The degeneracy between states $\ket{\pm}$ is lifted with a $B_{NV} = 1.1$\,mT magnetic field applied parallel to the chip and along the $[1,1,1]$ crystalline axis \footnote{note that $B_{NV}$ is wrongly mentioned to be $1.4$\,mT in the original article~\cite{Kubo.PhysRevLett.107.220501(2011),Kubo.PhysRevB.86.064514(2012)}}. The NV frequencies being sensitive only to the projection of $B_{NV}$ along the $N-V$ axis, two groups of NVs thus experience different Zeeman effects: those along $[1,1,1]$ (denoted $I$) and those along either of the three other $\left\langle 1,1,1\right\rangle $ axes (denoted $III$ as they are $3$ times more numerous). This results in four different ESR frequencies $\omega_{\pm I,\pm III}$. They are visible as $4$ avoided crossings in the resonator transmission spectrum (see Fig.~\ref{fig:NVQubit1}c) as a function of $\Phi$. From the data we also deduce the ensemble coupling constants $g_{\pm I}/2\pi = 2.9$\,MHz and $g_{\pm III}/2\pi=3.8$\,MHz. At a frequency much lower than the $4$ NV center ESR frequencies we observe another avoided crossing, this time due to the interaction with the transmon qubit. The transmission spectrum in Fig.~\ref{fig:NVQubit1}c shows that the tunable resonator can be used as a bus to dynamically transfer excitations from the qubit into the NV ensemble at one of its resonance frequencies, and vice-versa.

\begin{figure}[t!]
  \centering
  \includegraphics[width=16cm]{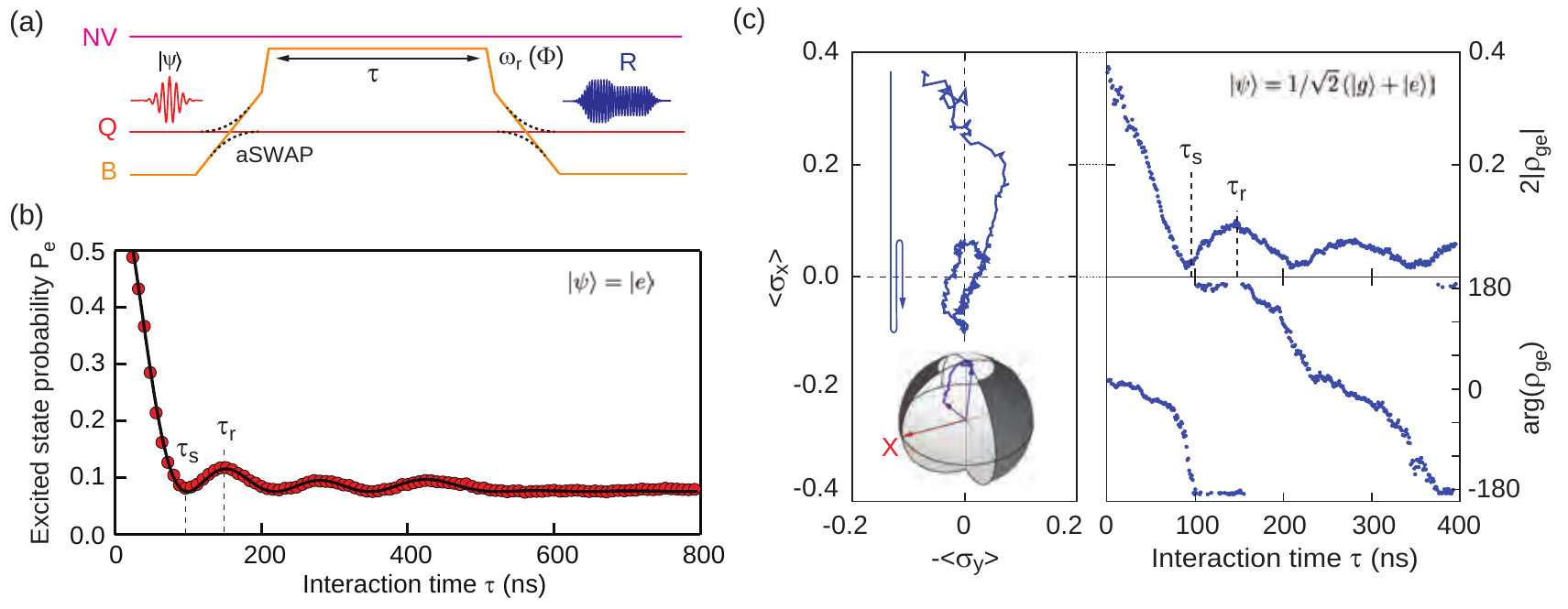}
  \caption{Storage of a qubit state into the spin ensemble. (a) Protocol. The qubit is prepared in state $\ket{\psi}$ with the appropriate pulse sequence. The state is then transferred to $\mathrm{B}$ by an adiabatic swap operation aSWAP, after which the bus is tuned near the spin resonance at frequency $\omega_\mathrm{B}(\Phi)$ during a time $\tau$. The final bus quantum state is transferred back to the qubit and either measured with a simple readout pulse yielding the excited state probability $P_e$, or with full quantum state tomography yielding the expectation values  $\langle \sigma_Z \rangle$, $\langle \sigma_X \rangle$, and $\langle \sigma_Y \rangle$ of the qubit state. (a) Qubit excited state probability $P_e(\tau)$ for $\omega_\mathrm{B}(\Phi) = \omega_{-I}$ and $\ket{\psi} = \ket{e}$. A single microwave photon is transferred into the spin ensemble at time $\tau_{s}$ and retrieved at time $\tau_r$. (b) Qubit quantum state tomography for $\omega_\mathrm{B}(\Phi) = \omega_{-I}$ and $\ket{\psi} = (\ket{g} + \ket{e})/\sqrt{2}$. (extracted from~\cite{Kubo.PhysRevLett.107.220501(2011)}).}
	\label{fig:NVQubit2}
	\end{figure}

The pulse sequence used to transfer an arbitrary qubit state $\ket{\psi}$ into the spin ensemble is shown in Fig~\ref{fig:NVQubit2}. The state is first transferred from the qubit into the bus resonator with an adiabatic SWAP gate (aSWAP operation), by adiabatically sweeping $\omega_\mathrm{B}$ across $\omega_\mathrm{Q}$. $\mathrm{B}$ is then brought at a frequency $\omega_\mathrm{B}(\Phi)$ in or near one of the spin ensemble resonances for a duration $\tau$;
the resulting $\mathrm{B}$ state is then transferred back into the qubit, which is finally read-out.

The result is shown in Fig.~\ref{fig:NVQubit2}, for $\omega_\mathrm{B}(\Phi) = \omega_{-I}$ and two different initial qubit states. For $\ket{\psi} = \ket{e}$ (see Fig.~\ref{fig:NVQubit2}a), an oscillation of small amplitude in $P_{e}$ is
observed, revealing the storage in the spin ensemble of the single quantum
of excitation initially in the qubit at $\tau_{s}=97$~ns, and its retrieval back into the qubit at $\tau_{r}=146$~ns. The relatively low value of $P_{e}(\tau_{r})$ is not only due to the spin ensemble inhomogeneous broadening, but also to an interference effect
caused by the HF structure of NV centers, as evidenced by the non-exponential
damping. The measurements are accurately
reproduced by a full calculation of the spin-resonator dynamics~\cite{Kurucz.PhysRevA.83.053852(2011),Diniz.PhysRevA.84.063810(2011),Kubo.PhysRevA.85.012333}
taking into account this HF structure, with the linewidth of each
HF peak as the only adjustable parameter. A linewidth of $1.6$~MHz (compatible with ODMR measurements shown in Fig.~\ref{fig:HybridsNV}d) is in this way determined for the spins in group $I$.

A quantum memory should be able not only to store states $\ket{0}$ and $\ket{1}$, but also an arbitrary superposition of these states with a well-defined phase. To demonstrate that, we repeat the experiment with $\ket{\psi} = (\ket{g} + \ket{e})/\sqrt{2}$ as shown in Fig.~\ref{fig:NVQubit2}b. For this experiment, we perform a full quantum tomography of the qubit state to test if the phase of the initial superposition state is preserved during the transfer to the spins. After substracting a trivial rotation
around $Z$ occurring at frequency $\left(\omega_{-I}-\omega_\mathrm{Q}\right)$,
we reconstruct the trajectory of this Bloch vector as a function of
the interaction time $\tau$. It is plotted in Fig. \ref{fig:NVQubit2}, together
with the off-diagonal element $\rho_{ge}$ of the final qubit density
matrix, which quantifies its coherence. We find that no coherence
is left in the qubit at the end of the sequence for $\tau=\tau_{s,I}$,
as expected for a full storage of the initial state into the ensemble.
Then, coherence is retrieved at $\tau=\tau_{r}$, although with
an amplitude $\sim5$ times smaller than its value at $\tau=0$ (i.e.
without interaction with the spins). Note the $\pi$ phase shift occurring
after each storage-retrieval cycle, characteristic of $2\pi$ rotations
in the two-level space $\left\{ \left|1_{B},0_{-I}\right\rangle ,\left|0_{B},1_{-I}\right\rangle \right\} $.
The combination of the results of Figs.~\ref{fig:NVQubit2}a and b
demonstrates that arbitrary superpositions of the two qubit states can be stored
and retrieved in a spin ensemble and thus represents a first proof-of-concept of a spin-based quantum memory
for superconducting qubits~\cite{Kubo.PhysRevLett.107.220501(2011)}. Similar results were obtained with a NV ensemble directly coupled to a flux-qubit~\cite{zhu_coherent_2011,SaitoPRL111-107008(2013)}, and with a magnonic excitation in YIG coupled to a transmon qubit via a three-dimensional metallic cavity~\cite{Tabuchi.Science.349.405(2015)}.

We now discuss the significance of these measurements in light of the quantum memory protocol described earlier. At first sight, one could conclude that the transfer efficiency of the qubit excitation into the spin memory is rather low, since we find $P_e(\tau_r)$ to be much lower than $P_e(0)$. This would be the case if the goal of the experiment was to simply use the bright mode as storage medium for the quantum state. But we consider the leakage of the quantum state from the bright mode into the bath of dark modes as a part of the whole memory protocol instead of a detrimental effect, as long as we can later retrieve this quantum information into the bright mode using refocusing pulses. In that perspective, the only loss of quantum information in the process comes from energy damping in the resonator (with a time constant $\kappa^{-1} = 3.3$\,$\mu$s), and from non-static spin dephasing (with time constant $T_2 = 7.3$\,$\mu$s). An estimate of the fidelity is thus given by $\exp(- (\kappa + T_2^{-1}) \tau_s / 2) \approx 95 \%$, which is sufficiently high for demonstrating a quantum memory and could be optimized with improved parameters. Note that although a single photon can be fully transferred from the bus resonator into the spins in a time $\tau_s$ of the order or shorter than $T_2^*$, the spin ensemble memory would still need a time of order $2-3 \times T_2^*$ for the bright mode to be damped into the dark modes as shown in Fig.~\ref{fig:NVQubit2}, after which a new incoming quantum state can be stored.

\section{Storage and retrieval of weak microwave pulses}
\label{sec:StorageRetrieval}

\begin{figure}[t!]
  \centering
  \includegraphics[width=11cm]{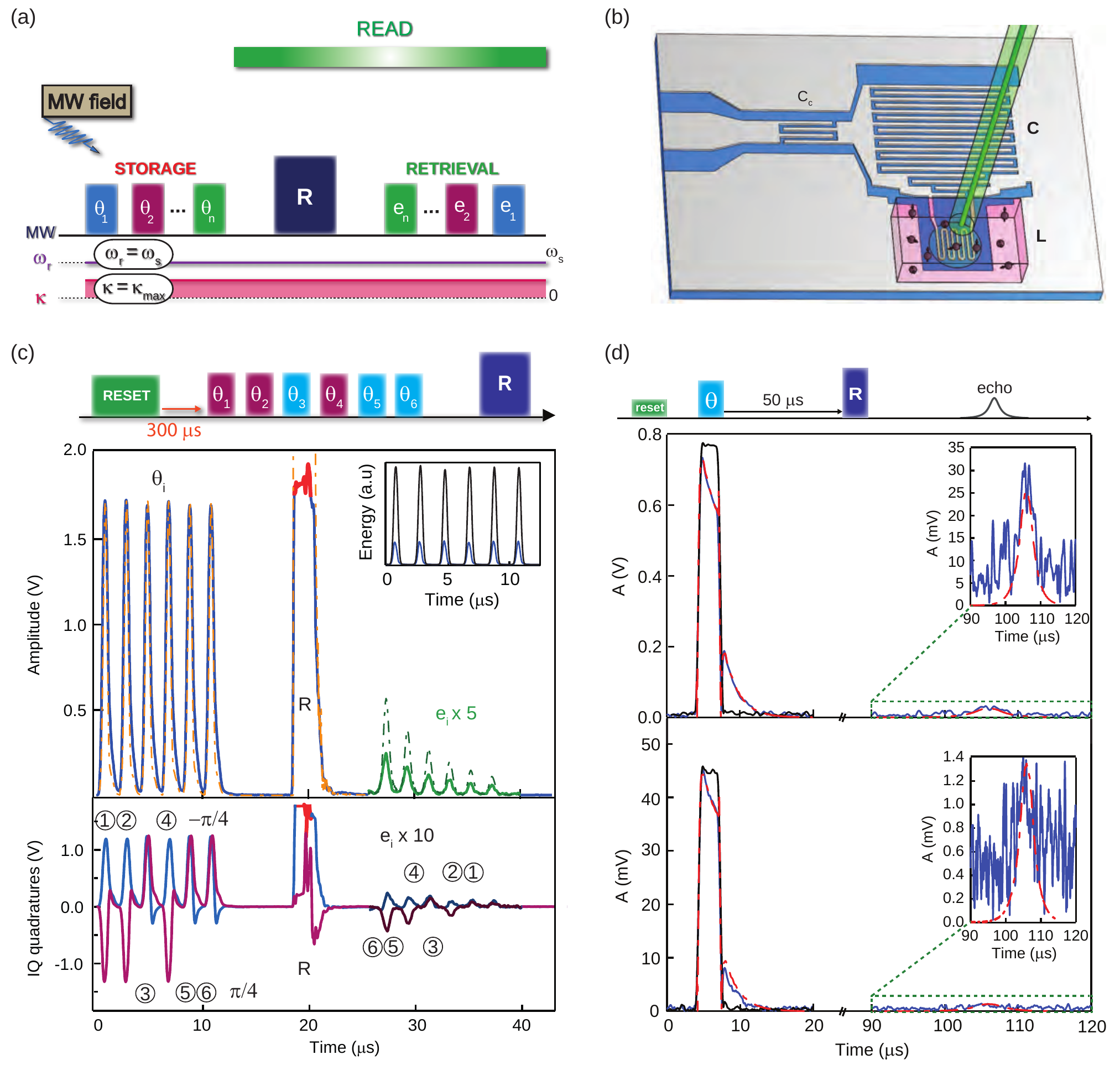}
  \caption{Retrieval of few-photon fields stored in a spin ensemble. (a) Scheme of the multimode two-pulse echo (2PE) protocol. Successive few-photon microwave pulses $\theta_i$ are stored in the spin ensemble. A single refocusing pulse $R$ acts as time-reversal for the spins and triggers the retrieval of the stored pulses as echoes $e_i$ in reverse order. (b) Schematic view of the hybrid circuit. The ensemble of NV center spins (in pink) is coupled to a planar superconducting lumped element resonator. Laser pulses are applied to the spin ensemble through an optical fiber glued to the top face of the diamond. (c) Six consecutive microwave pulses $\theta_i$ with a varying phase and identical amplitude corresponding to $\approx 10^4$ photons in the resonator are absorbed by the spin ensemble, followed $10$\,$\mu$s later by a strong refocusing pulse $R$ applied to trigger their retrieval. The output signal shows the reflected pulses $\theta_i$ (after partial absorption by the spins) and $R$ (its amplitude being trimmed by amplifier saturation, shown in red), as well as the six reemitted echoes $e_i$ (magnified by a factor of 5). Inset: Comparison between the energies of the reflected $\theta_i$ pulses with the spins saturated (black line) or reset in their ground state (blue line) shows that about $75$\% of the incident power is absorbed by the spins. (d) Storage and retrieval at the single photon level in an isotopically enriched diamond sample. The echo (e) is obtained for a low power incoming $\theta$ pulse populating the resonator with only $\sim 60$ photons (top) and $<1$ photon (bottom) with $0.3$\,\% efficiency after $100$\,$\mu$s storage. Dashed lines in c and d are the result of numerical simulations (see main text). (extracted from~\cite{Grezes.PHYSREVX4.021049(2014),Grezes.PhysRevA.92.020301(2015)})}
	\label{fig:MultimodeStorage}
\end{figure}

The main obstacle to the implementation of the {\it read} step of the memory protocol is to extend refocusing techniques such as the Hahn echo down to the quantum regime, in a hybrid quantum circuit. For that, we test an essential building block of the quantum memory protocol of paragraph~\ref{sec:protocol}, known as the Two-Pulse Echo (2PE) and shown schematically in Fig.~\ref{fig:MultimodeStorage}a. The 2PE consists in storing weak pulses $\theta_i$ containing only a few microwave photons into the spin ensemble at times $t_i$, and applying a $\pi$ pulse at time $\tau$ which triggers the refocusing of the spins at times $2 \tau - t_i$ (that is, in reverse order) and thus the emission of echoes $e_i$ in the detection waveguide~\cite{anderson:JApplPhys.26.1324(1955)}. 

The storage of successive high-power microwave pulses into a spin ensemble has already been demonstrated\cite{anderson:JApplPhys.26.1324(1955),Wu.PhysRevLett.105.140503(2010),Probst.PhysRevB.92.014421(2015)}; one experimental difficulty to reach the few-photon regime is that the spin memory now needs to be actively reset between consecutive sequences. Indeed, whereas in the experiment described in paragraph~\ref{sec:NVQubit} at most one energy quantum is transferred to the spin ensemble in each sequence, which is negligible compared to the average thermal excitation of the spins (approximately $5\%$ at $50$\,mK), pulses in a Hahn echo sequences excite all of the $10^{11}$ spins, which need to come back to thermal equilibrium before the next experiment can be performed. At low temperatures, spin relaxation processes are strongly suppressed~\cite{Grezes.PHYSREVX4.021049(2014)}, leading to prohibitively low repetition rates unless an active reset is implemented. 

The device is sketched in Fig.~\ref{fig:MultimodeStorage}b~\cite{Grezes.PHYSREVX4.021049(2014)}. It consists of a superconducting planar LC resonator of frequency $\omega_r / 2 \pi = 2.88$\,GHz, coupled inductively to a diamond crystal of the HPHT type containing the NV center ensemble with $\rho \approx 2$\,ppm. The spins are resonant with $\omega_r$ even without applying a magnetic field. The resonator is capacitively coupled to a measuring line, through which microwave pulses can be sent, yielding a quality factor $Q=80$ chosen on purpose so that $g_{ens} < \kappa$, in order to avoid superradiant relaxation of the spin ensemble after the refocusing pulse~\cite{Julsgaard.PhysRevA.85.013844(2012)}. The microwave signal reflected on the resonator is amplified, and its amplitude and phase measured in a homodyne detection setup at room-temperature. An optical fiber, glued to the diamond crystal on top of the resonator inductance, makes it possible to send pulses of laser light at $532$\,nm to optically reset the NV centers into their ground state. Optical pulses of $1.5$\,mW power and $1$\,s duration are sufficient to repump $90\%$ of the NVs into their ground state~\cite{Grezes.PHYSREVX4.021049(2014),Grezes.PhD(2014)}, although this causes the cryostat temperature to raise up to $400$\,mK.

Using the active reset makes it possible to repeat the experimental sequences at a $1$\,Hz rate, and therefore to measure the 2PE at very low powers. As shown in Fig.~\ref{fig:MultimodeStorage}, six successive microwave pulses $\theta_i$ of frequency $\approx \omega_r$ and containing each $10^4$ photons, are absorbed by the NV ensemble. At a later time, a strong microwave pulse performs the refocusing operation. Due to the geometry of the experiment, the microwave magnetic field generated by the resonator is inhomogeneous spatially; as a result it is not possible to apply a $\pi$ pulse to all the spins, since for a given input power they experience different Rabi frequencies depending on their position in the crystal. This reduces the echo amplitude but does not prevent from measuring it. After averaging the signal over $10^4$ identical sequences, the expected six echo pulses are indeed observed in the reflected signal, with an intensity corresponding to $5$ photons on average in the resonator. Successive spin-echoes were also observed in an ensemble of rare-earth ions in a silicate crystal coupled to a coplanar waveguide resonator~\cite{Probst.PhysRevB.92.014421(2015)}

One quantity of interest is the ratio of the energy emitted during the echo over the energy of the incoming pulse, named "echo efficiency" $E$; in the perspective of an efficient quantum memory it would be desirable to observe $E$ as close as possible to $1$. Our experiment, however, reaches only $E \approx 2 \times 10^{-4}$. Although this value is far from sufficient for a future quantum memory, it already constitutes a significant improvement over the state-of-the-art~\cite{Wu.PhysRevLett.105.140503(2010)} where $E \approx 10^{-10}$. To understand the finite echo efficiency, we have performed detailed numerical simulations, taking into account the measured inhomogeneous broadening of the NV transition, its hyperfine structure, the spatial inhomogeneity of the resonator microwave field, and the measured spin coherence time $T_2 \approx 8 \mu \mathrm{s}$, as explained in more details in\cite{Julsgaard.PhysRevLett.110.250503,Grezes.PHYSREVX4.021049(2014)}. The result is in good agreement with the data as shown in Fig.~\ref{fig:MultimodeStorage}b, which shows that the main limitation of the retrieval efficiency in this experiment is the finite NV coherence time due to the large concentration of P1 centers ($\approx 16$\,ppm) in this sample. 

In more recent experiments~\cite{Grezes.PhysRevA.92.020301(2015)}, a diamond sample having a reduced concentration of P1 centers as well as isotopic enrichment in the nuclear-spin-free $^{12}C$ is used. The longer coherence time ($T_2 = 84 \mu \mathrm{s}$) makes it possible to achieve $3.10^{-3}$ retrieval efficiency after $100$\,$\mu$s (see Fig.~\ref{fig:MultimodeStorage}c), a one order of magnitude improvement in both storage time and retrieval efficiency compared to the previous results. The storage pulse contains $< 1$ photon, demonstrating that when the retrieval efficiency is sufficient, it is indeed possible to detect an echo down to the single-photon level as required for the memory protocol.

\section{Conclusion and perspectives}
\label{sec:conclusion}

We have presented a memory protocol able to store and retrieve on-demand the state of a large number of qubits in a spin ensemble and we demonstrated building blocks of its implementation with NV centers in diamond. The protocol requires a challenging combination of the most advanced techniques of superconducting quantum circuits and pulsed electron spin resonance. The first step, the coherent storage of a qubit state into the spin ensemble, was achieved in a first experiment with $0.95$ efficiency. The second step, the retrieval using refocusing techniques of microwave pulses stored in the spin ensemble, was demonstrated with a coherent state input at the single-photon level with $\sim 3 \cdot 10^{-3}$ efficiency after $100$\,$\mu$s memory time. The reset of the spin memory by optical repumping was adapted in a dilution refrigerator. Finally the multi-mode character of the NV spin ensemble memory was evidenced in a multiple microwave pulses storage and retrieval experiment. 

Having presented these "`building block"' experiments, it is interesting to envision the future directions of this research project. 

One clear drawback of the experiments discussed in paragraph~\ref{sec:StorageRetrieval} is the spread of Rabi frequencies over the whole spin ensemble, which forbids to perform well-defined Rabi rotations, due to the spatial inhomogeneity of the microwave field generated by the planar resonator. This may be solved by implanting the spins at well-defined locations of a substrate, on top of which the resonator can be subsequently patterned in such a way that the oscillating magnetic field is homogeneous over the whole spin ensemble location. In a related recent experiment with donors in silicon~\cite{bienfait2015reaching} coupled to a superconducting planar micro-resonator, mechanical strain exerted by the resonator thin-film has been shown to effectively select the frequency of spins located just below the superconducting inductance, which allowed to perform well-defined Rabi rotations. 

Although $T_2=100 \mu \mathrm{s}$ is already sufficient for first demonstrations, the echo efficiency would also be improved if the spin coherence time was longer. Interestingly however, an important contribution to decoherence is the unavoidable dipole-dipole interaction between NV centers~\cite{Grezes.PhysRevA.92.020301(2015)}. A reduced spin concentration would therefore increase further $T_2$, but would also necessitate a larger resonator quality factor to reach strong coupling. Another interesting option could be to incorporate in the memory protocol pulse sequences designed to refocus dipolar interactions~\cite{Waugh.PhysRevLett.20.180(1968)}. A last promising direction is to rely on so-called clock transitions, where the spin frequency is stationary respectively to the applied magnetic field, which suppresses the dephasing due to dipole-dipole interactions and leads to orders of magnitude gains in coherence time even for large spin concentration. Such clock transitions are found in bismuth donors in silicon~\cite{Wolfowicz.NatureNano.8.561(2013)}, rare-earth ions in YSO crystals\cite{Zhong.Nature.517.177(2015)}, but also in NV centers in diamond~\cite{Dolde2011}.

Another aspect that was not discussed here is the need to use a resonator with dynamically tunable frequency and quality factor, in particular to silence the first echo. The tunable resonator design used in the experiments reported in paragraphs $4$ and $5$ is not directly compatible with the high-power microwave pulses used in our spin-echo experiments, since the microwave currents would then overcome the critical currents of the Josephson junctions used as tuning elements. Nevertheless, a  combination of coupled linear and tunable resonators may be employed, as discussed in more details in~\cite{Grezes.PhD(2014)}.

Finally, it would be of high interest to measure not only the average amplitude of the microwave signals emitted by the spins, but also to perform their complete quantum tomography, in order to see if the spins add extra noise to the input signal. This could be achieved by introducing a Josephson parametric amplifier in the output line, which is known to operate close to the quantum limit thus enabling to directly measure the quantum statistics of the spin-echo signal and to measure the fidelity of our quantum memory~\cite{Julsgaard.PhysRevA.89.012333(2014)}.

Combining these improvements with the results reported in this article makes us confident that the realization of an operational quantum memory able to store hundreds of qubit states over seconds is within reach.

{\bf Acknowledgements} We acknowledge technical support from P. S{\'e}nat, D. Duet, J.-C. Tack, P. Pari, P. Forget, as well as useful discussions within the Quantronics group and with A. Dr{\'e}au, T. Chaneli{\`e}re and J. Morton. We acknowledge support of the French National Research Agency (ANR) with the QINVC project from CHISTERA program, of the European project SCALEQIT, of the C'Nano IdF project QUANTROCRYO, of JST, and of JSPS KAKENHI (no. 26246001). Y. Kubo acknowledges support from the JSPS, and B. Julsgaard and  K. M{\o}lmer from the Villum Foundation.

\end{document}